\documentclass{emulateapj}

\shorttitle{GRB080319B}
\shortauthors{D'Elia et al.}

\begin{document}
%\title{The High Resolution View of the Brightest GRB Optical Afterglow
%Ever Recorded: GRB080319B
\title{The prompt, high resolution spectroscopic view of the
``naked-eye'' GRB080319B
\footnote{Based on observations collected at 
the European Southern Observatory (ESO) with the VLT/Kueyen telescope, 
Paranal, Chile, in the framework of program 080.A-0398.}}
\author{V. D'Elia\altaffilmark{1}, 
F. Fiore\altaffilmark{1}, R. Perna\altaffilmark{2}, Y. Krongold\altaffilmark{3}, 
S. Covino\altaffilmark{4},  D. Fugazza\altaffilmark{4}, D. Lazzati\altaffilmark{2}, 
F. Nicastro\altaffilmark{1}, L.A. Antonelli\altaffilmark{1}, S. Campana\altaffilmark{4}, 
G. Chincarini\altaffilmark{4,5}, P. D'Avanzo\altaffilmark{4},  M. Della Valle\altaffilmark{6}, 
P. Goldoni\altaffilmark{7}, D. Guetta\altaffilmark{1}, C. Guidorzi\altaffilmark{4},
E.J.A. Meurs\altaffilmark{8,9}, F. Mirabel\altaffilmark{10}, E. Molinari\altaffilmark{4}, 
L. Norci\altaffilmark{9}, S. Piranomonte\altaffilmark{1}, L. Stella\altaffilmark{1}, 
G. Stratta\altaffilmark{11}, G. Tagliaferri\altaffilmark{4}, P. Ward\altaffilmark{8,12}.
}

\altaffiltext{1}{INAF, Osservatorio Astronomico di Roma, 
via Frascati 33, Monteporzio (Rm), I00040, Italy;}
\altaffiltext{2}{JILA and Department of Astrophysical and Planetary Science, 
CU Boulder, Boulder, 80309, USA;}
\altaffiltext{3}{Instituto de Astronomia, Universidad Nacional Autonomica de 
Mexico, Apartado Postal 70-264, 04510 Mexico DF;}
\altaffiltext{4}{INAF, Osservatorio Astronomico di Brera, via E. Bianchi 46, 23807 
Merate (LC), Italy;}
\altaffiltext{5}{Universita' di Milano Bicocca, piazza della Scienza 3, 
20126 Milano, Italy}
\altaffiltext{6}{INAF, Osservatorio Astronomico di Capodimonte, salita Moiarello 16, Napoli, Italy;}
\altaffiltext{7}{APC/UMR 7164, Service dÕAstrophysique, CEA Centre de Saclay;}
\altaffiltext{8}{School of Cosmic Physics, DIAS, 31 Fitzwilliam Place, Dublin 4, Ireland;}
\altaffiltext{9}{School of Physical Sciences and NCPST, DCU, Glasnevin, Dublin 9, Ireland;}
\altaffiltext{10}{European Southern Observatory, Casilla 19001, Santiago, Chile;}
\altaffiltext{11}{ASI Science Data Center, via Galileo Galilei, 
00044 Frascati, Italy}
\altaffiltext{12}{Mullard Space Science Laboratory, Dorking, Surrey RH5 6NT, UK.}

\email{delia@oa-roma.inaf.it}

\begin{abstract}

GRB080319B reached 5th optical magnitude during the burst prompt
emission. Thanks to the VLT/UVES rapid response mode, we observed its
afterglow just 8m:30s after the GRB onset when the magnitude was
R $\sim 12$. This allowed us to obtain the best signal-to-noise, high
resolution spectrum of a GRB afterglow ever (S/N per resolution
element $\sim50$). The spectrum is rich of absorption features
belonging to the main system at z=0.937, divided in at least six
components spanning a total velocity range of 100 km s$^{-1}$. The
VLT/UVES observations caught the absorbing gas in a highly excited
state, producing the strongest \ion{Fe}{2} fine structure lines ever
observed in a GRB.  A few hours later the optical depth of these lines
was reduced by a factor of 4-20, and the optical/UV flux by a factor
of $\sim60$. This proves that the excitation of the observed fine
structure lines is due to ``pumping'' by the GRB UV photons. A
comparison of the observed ratio between the number of photons
absorbed by the excited state and those in the \ion{Fe}{2} ground
state suggests that the six absorbers are $\sim 2-6$ kpc from the GRB
site, with component I $\sim3$ times closer to the GRB site than
components III to VI. Component I is characterized also by the lack of
\ion{Mg}{1} absorption, unlike all other components.  This may be due
both to a closer distance and a lower density, suggesting a structured
ISM in this galaxy complex.

\end{abstract}

\section{Introduction}

For a few hours after their onset, Gamma Ray Burst (GRB) afterglows
are the brightest beacons in the far Universe. In a small fraction of
the cases, extremely bright optical transient emission is associated with
the GRB event, offering a superb opportunity to investigate high--z
galaxies through high resolution spectroscopy of the optical
transient. The study of the rich absorption spectra can yield unique
information on the gas in the GRB environment and  the physical,
chemical and dynamical state and geometry of the inter--stellar matter
(ISM) of intervening galaxies, including the GRB host galaxy.

GRB080319B was discovered by the Burst Alert Telescope (BAT)
instrument on board Swift on 2008, March 19, at 06:12:49 UT. Swift
slewed to the target in less than 1 minute and a bright afterglow was
found by both the X-Ray Telescope (XRT) and UV-Optical Telescope
(UVOT) at RA = 14h 31m 40.7s, Dec = +36$^o$ $18'$ $14.7"$ (Racusin et
al. 2008a) with observations starting 60.5 and 175 s after the
trigger, respectively. The field of GRB080319B was imaged by the "Pi
of the Sky" apparatus located at Las Campanas Observatory before,
during and after the GRB event (Cwiok et al. 2008). The field was also
targetted by the robotic telescope REM just 43 s after the BAT trigger
(Covino et al. 2008a, b). The TORTORA wide-field optical camera (12 cm
diameter, 20$\times$25 deg FOV, TV-CCD, unfiltered) mounted on REM also
imaged the field before, during and after the GRB event with good
temporal resolution (Karpov et al. 2008).  These observations show
that the GRB reached the magnitudes $V = 5.3 $ about $20$ s and $H =
4.2$ about $50$ s after the trigger.  This makes GRB080319B the
brightest GRB ever recorded at optical wavelengths (Bloom et al. 2008,
Racusin et al. 2008b).

\begin{table*}[ht]
\caption{\bf GRB080319B journal of observations}
{\footnotesize
\smallskip
\label{obs_log}
\begin{tabular}{lccccccc}
\hline
\hline
Obs  & UT observation & T. from burst (s) & Exp. (s) & S/N range & Dichroics & Arms & R mag \\
\hline
RRM 1  & 2008 Mar 19, 06:21:26 & 517   & 600  & $30 \div 50$ & 2     & Blue + Red & $12 \div 13$\\
RRM 2  & 2008 Mar 19, 08:06:42 & 6833  & 1800 & $7  \div 12$ & 1 + 2 & Blue + Red & $16 \div 17$\\
ToO    & 2008 Mar 19, 09:07:22 & 10482 & 1200 & $5  \div 8 $ & 1 + 2 & Blue + Red & $16 \div 17$\\

\hline
\end{tabular}
}
\end{table*}

The optical afterglow of GRB080319B was observed at high resolution
with VLT/UVES starting just 8m:30s after the BAT trigger, thanks to
the VLT rapid response mode (RRM), when its magnitude was R$\sim12
\div 13$. This allowed us to obtain the best signal-to-noise, high
resolution spectrum of a GRB afterglow ever (S/N per resolution
element $\sim50$). Two further RRM and target of opportunity (ToO)
observations were obtained $ 2 - 3 $ hours after the event.  Several
absorption systems are present in these spectra.  Vreeswijk et
al. (2008) identify the highest redshift system at 0.937 as the GRB
host galaxy.

This paper concentrates on the analysis of the \ion{Fe}{2} excited
lines associated with the main system at z=0.937 and on their
variability.  Section 2 describes the datasets and data analysis;
Section 3 presents the UVES spectroscopy and discusses the absorption
features and their variability; Section 4 concerns the evaluation of
the distance of the absorbers from the GRB explosion site; our
conclusions are given in Section 5. A $H_0=70$ km s$^{-1}$ Mpc$^{-1}$,
$\Omega_M$=0.3, $\Omega_{\Lambda}=0.7$ cosmology is adopted
throughout.

\begin{figure*}[ht]
\centering
\begin{tabular}{cc}
\includegraphics[width=6cm, angle=-90]{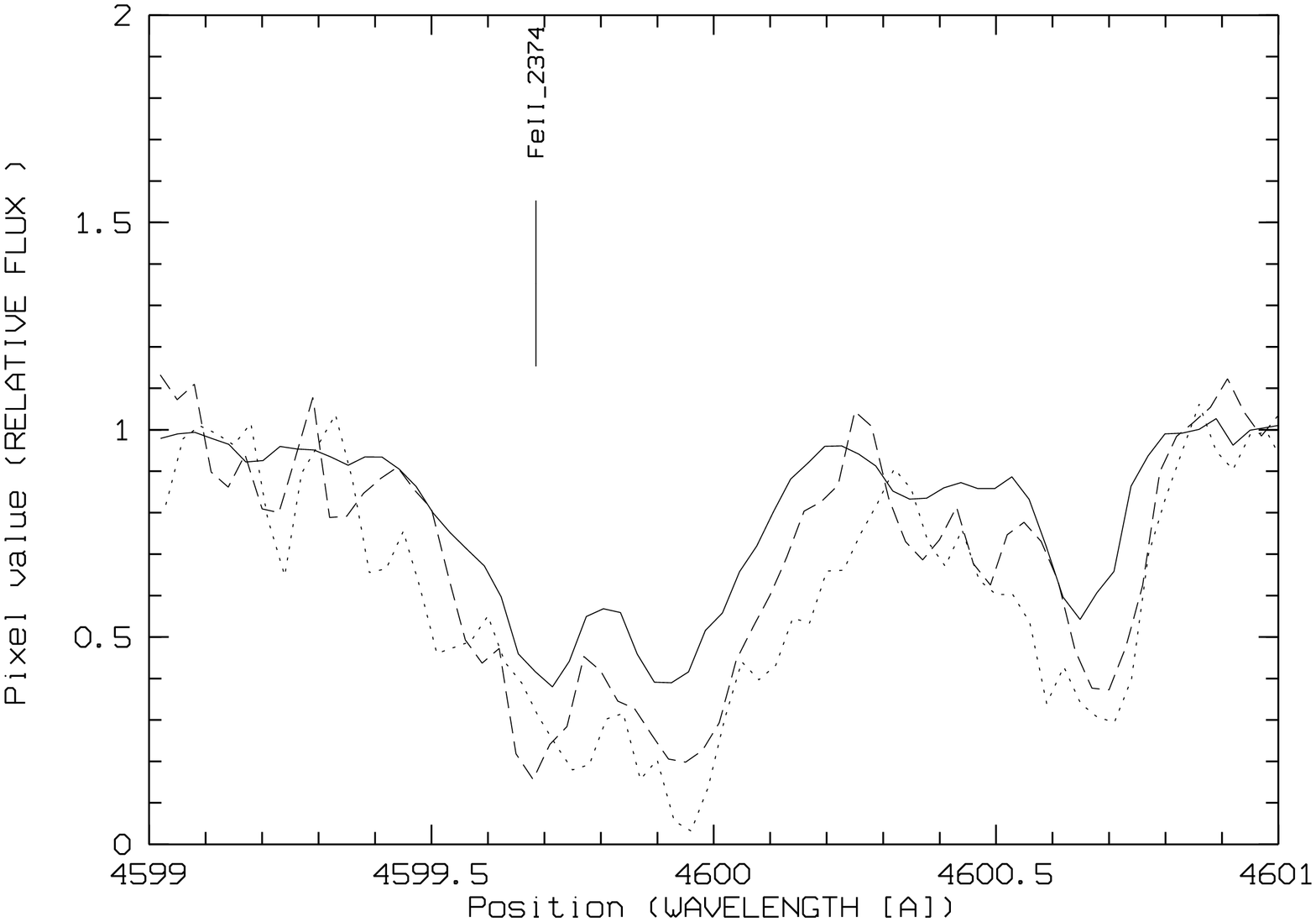}
\includegraphics[width=6cm, angle=-90]{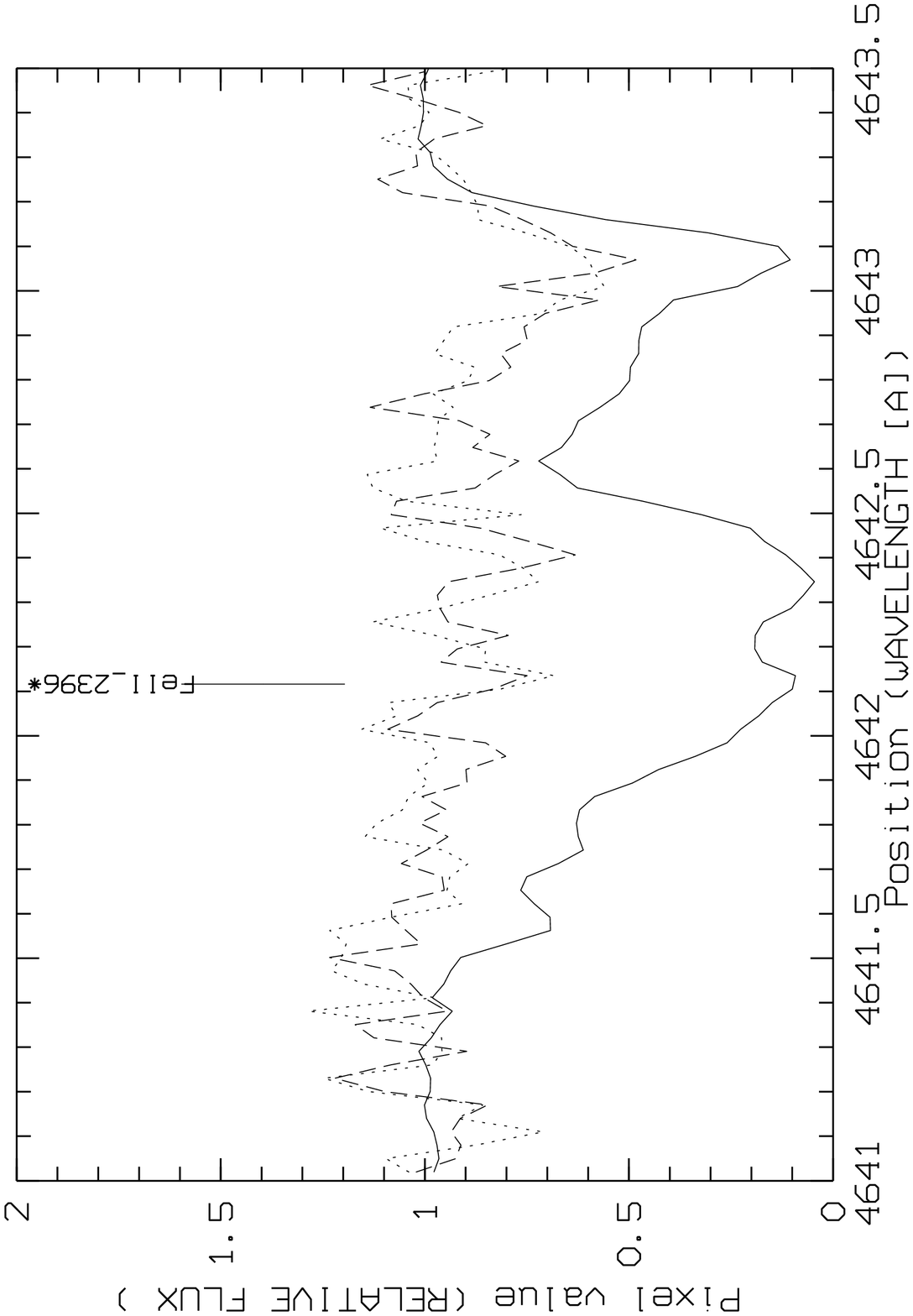}
\end{tabular}
\caption{ The UVES spectra of GRB080319B around the
\ion{Fe}{2}$\lambda$2374 (left panel) and \ion{Fe}{2}$\lambda$2396*
(right panel) transitions. Solid lines refer to the first epoch
spectrum (8m30s after the Swift trigger), dashed lines to the second
epoch spectrum (1.9 hours after the GRB event), and dotted lines to
the the third epoch spectrum (2.9 hours after the GRB event).}
\label{spettri1}
\end{figure*}

\section{Observations and data analysis}

We observed the bright afterglow of GRB080319B in the framework of the
ESO program 080.A-0398 with the VLT/UVES (Dekker et al. 2000).
The Observation Log is reported in Table \ref{obs_log}.  Both UVES
dichroics, as well as the red and the blue arms, were used.

The first, 10min observation, was performed in RRM and started just
8m:30s after the GRB event, when the afterglow was extremely
bright (R=12-13). This afforded a S/N=$30\div 50$ per resolution
element. Two more UVES observations followed, the first one again in
RRM mode, activated in the framework of program 080.D-0526 and
starting 1.9 hours after the GRB event, and the second a ToO, starting
2.9 hours after the GRB, see Table \ref{obs_log}.

Data reduction was carried out by using the UVES pipeline
(Ballester et al. 2000). The final useful spectra extend from 
$\sim 3800$~\AA{} to $\sim 9500$~\AA.
%The total spectrum was rebinned to 0.1~\AA~ to increase the
%signal-to-noise ratio. 
The resolution element, set to two pixels, ranges then from 4
km\,s$^{-1}$ at 4500~\AA{} to 1.9 km\,s$^{-1}$ at 9000~\AA.  The noise
spectrum, used to determine the errors on the best fit line
parameters, was calculated from the real-background-subtracted spectra
using line-free regions. This takes into account both statistical
and systematic errors in the pipeline processing and background
subtraction. 

\section{UVES spectroscopy of excited lines}

The three UVES observations were analyzed in the MIDAS environment
using the {\sc fitlyman} procedure (Fontana \& Ballester 1995). The
highest z system present in these spectra is at z=0.937, as also
reported by Vreeswijk et al. (2008). This system presents absorption
features from the ground states of MgI, MgII, FeII and several FeII
fine structure lines (FeII* hereafter).  The most striking feature in
the UVES spectra is the variation of the opacity of the fine structure
lines between the first and the second UVES
observation. Fig. 1 shows the \ion{Fe}{2}$\lambda$2374
and \ion{Fe}{2}$^*\lambda$2396 absorption features in the three
epochs.  We see strong variations of both lines. While the strength of
the \ion{Fe}{2}$\lambda$2374 absorption increases from the first to
the third epoch, strong \ion{Fe}{2}$^*\lambda$2396 absorption is
present only in the first spectrum and nearly disappears in the second
and third spectra. The huge variations of \ion{Fe}{2} fine structure
lines imply that ``pumping'' by the GRB UV photons is the main
mechanism for populating the excited states (Silva \& Viegas 2002;
Prochaska et al. 2006; Vreeswijk et al. 2007).

\begin{figure}[h]
\centering
\begin{tabular}{cc}
\includegraphics[width=9cm,height=8.5cm,angle=-90]{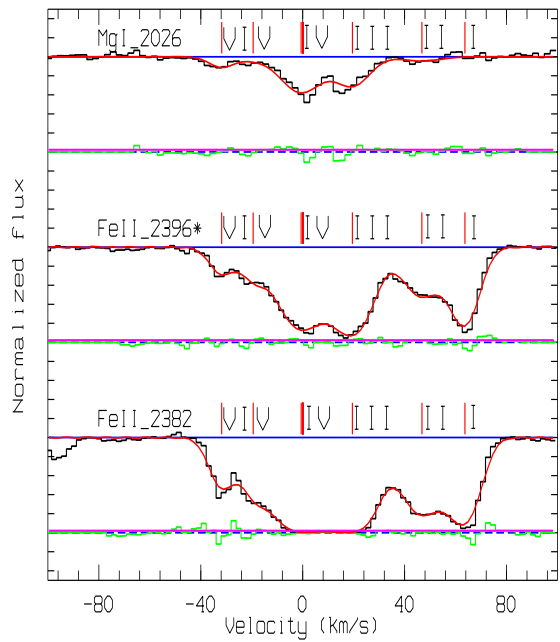}
\end{tabular}
\caption{ The first UVES spectrum of GRB080319B around the
MgI$\lambda$2026, FeII$\lambda$2396$^*$ and FeII$\lambda$2382
transitions. The solid line shows the six component fit (I to VI from
higher to lower redshift). The velocity position of the components is
marked with vertical lines, as well as the zero point at $z=0.9371$.}
\label{spettri2}
\end{figure}

UVES spectra of bright GRB afterglows have always revealed a complex
structure of the absorption system associated with the GRB host
galaxy, reflecting the clumpy nature of the ISM (see e.g. D'Elia et
al. 2007). This is confirmed by the UVES spectra of GRB080319B.  A
detailed line fitting was performed using a Voigt profile with three
parameters: the line wavelength, column density and Doppler parameter
$b$.  Several absorption features were fitted simultaneously by
keeping the redshift and $b$ value of each component fixed at their
common values (best fit $b$ values in the $3 \div 10$ range). The
\ion{Fe}{2}$^*\lambda$2396 absorption lines are not saturated, and can
be used to guide the identification of different components.
Statistically acceptable fits to the first epoch UVES spectrum are
obtained by using six components. These span a range of $\sim 100$
km s$^{-1}$ in velocity space.  Fig. \ref{spettri2} shows the best fitting
model to the \ion{Mg}{1}$\lambda$2026, \ion{Fe}{2}$\lambda$2382 and
\ion{Fe}{2}$^*\lambda$2396 lines. The lower S/N spectra from the
second and third epochs were then fitted by fixing the z and $b$
parameters of each component at their respective best fit values found
for the first epoch, highest S/N spectrum.  

Table 2 gives the \ion{Mg}{1} and \ion{Fe}{2} and column densities of
each of the six components in the three epochs. Components are labeled
from I to VI for decreasing wavelengths (and decreasing redshift, or
positive velocity shift with respect to a zero point, placed at
z=0.9371). \ion{Fe}{2} is represented by the ground, first excited
($4F$) and second excited ($^4D$) levels. Fine structures of each
level are marked with asterisks; the ground state shows four fine
structure levels, the excited ones just the first level. The second
column indicates which transitions have been used to evaluate the
column density of each ionic specie.  Strong \ion{Mg}{2} absorption is
present for all components, but reliable column densities cannot be
derived for this ion because the lines are strongly saturated.  The
column density uncertainties are given at the $1\sigma$ confidence
level, while upper limits are at a 90\% confidence level
(i.e. 1.6$\sigma$).  The column densities derived from the second
epoch spectrum are always consistent with those derived from the third
epoch spectrum, to within their relatively large errors. Thus, in
order to improve the $S/N$, we also added together the second and
third epoch spectra and repeated the fits.

\ion{Mg}{1} is detected for all components but I. The \ion{Mg}{1}
column density of the five detected components is consistent with a
constant value (within each component) at all epochs.  Conversely, we
see strong variations in time of both \ion{Fe}{2} excited and ground
state lines for all six components.  The \ion{Fe}{2} fine
structures line of the lower redshift components underwent the
strongest variations, as most of these lines are not detected in the
second and third epoch spectra. The \ion{Fe}{2} first fine structure line
of the highest redshift component I varies less, and it is still
detected in the second and third epoch spectra. Fig. 3 compares the
column density of the \ion{Fe}{2}$^*\lambda$2396 line of the six
components in the first epoch spectrum to that measured 2-3 hr
later. The column density of component I dropped by a factor of
$\sim4$, while that of component III dropped by a factor of $\sim20$
(Table 3).  On the other hand, the column density of ground state
\ion{Fe}{2} increased by a factor of 1.3-2 for all six components
(Table 3).  The de-excitation of the excited levels into ground state
levels, as time passes and the UV radiation field diminishes, is
certainly contributing to this increase.  For all components, the
increase in the column density of the \ion{Fe}{2} resonant line is
consistent with the decrease of the excited lines within $1
\sigma$. This is a first indication that the absorbing medium must be
relatively distant, since photoionization of the medium by the burst
photons, predicted to be important in the vicinity of the source
(Perna \& Loeb 1998; Perna \& Lazzati 2002) appears to be negligible
here.

\begin{figure}[h]
\centering
\begin{tabular}{cc}
\includegraphics[width=8.5cm, angle=0]{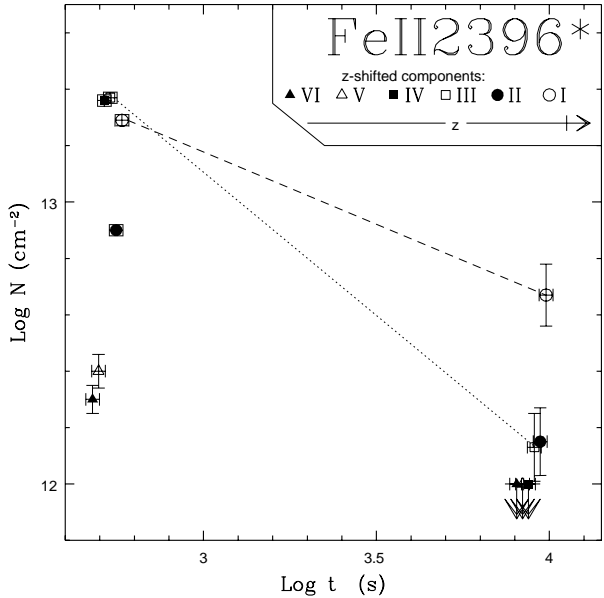}
\end{tabular}
\caption{ The column density of the FeII$\lambda$2396$^*$ line for the
six components as a function of time. For clarity reasons, components
have been slightly shifted with each other. Late time points represent
the observations 2 and 3 added together.  Note that the highest
redshift component I varies less than the lower redshift components
III and IV (the dashed and dotted lines are for components I and III,
respectively).\ }
\label{var}
\end{figure}

\begin{table*}[ht]
\caption{\bf MgI, FeII and FeII* column densities for the six components 
at three epochs.}
{\footnotesize
\smallskip
\begin{tabular}{lc|ccccccc}
\hline
\hline
Specie & Trans.   & Obs.  & I (64 km/s) & II (47 km/s)  & III (20 km/s) & IV (0 km/s) & V (-20 km/s)& VI (-32 km/s)\\
\hline
Mg I          &  $\lambda$2026  & 1& $ < 11.80      $  & $12.14 \pm 0.10$ &    $13.00 \pm  0.02  $ & $13.18 \pm 0.01   $  & $11.83 \pm  0.17  $ & $12.38  \pm 0.05  $ \\
              &  $\lambda$2852  & 2& $ < 11.2       $  & $12.09 \pm 0.03$ &    $13.06 \pm  0.08  $ & $12.94 \pm 0.12   $  & $11.77 \pm  0.06  $ & $12.02  \pm 0.05  $ \\
              &                 & 3& $ < 11.6       $  & $12.05 \pm 0.04$ &    $13.39 \pm  0.11  $ & $12.87 \pm 0.10   $  & $11.81 \pm  0.07  $ & $12.05  \pm 0.07  $ \\
              &                 &2+3&$ < 11.0       $  & $12.08 \pm 0.02$ &    $13.18 \pm  0.06  $ & $12.95 \pm 0.07   $  & $11.80 \pm  0.05  $ & $12.07  \pm 0.05  $ \\
\hline
FeII          &  $\lambda$2374  & 1& $13.52 \pm 0.01$  & $13.11 \pm 0.02$ &    $13.84 \pm  0.02  $ & $13.79 \pm 0.02   $  & $12.76 \pm  0.02  $ & $12.77  \pm 0.02  $ \\
              &  $\lambda$2382  & 2& $13.78 \pm 0.05$  & $13.26 \pm 0.09$ &    $14.13 \pm  0.05  $ & $14.01 \pm 0.06   $  & $13.11 \pm  0.06  $ & $12.86  \pm 0.22  $ \\
              &                 & 3& $13.99 \pm 0.07$  & $13.19 \pm 0.17$ &    $14.32 \pm  0.10  $ & $13.99 \pm 0.11   $  & $12.77 \pm  0.85  $ & $12.81  \pm 0.34  $ \\
              &                 &2+3&$13.87 \pm 0.04$  & $13.24 \pm 0.10$ &    $14.19 \pm  0.08  $ & $14.00 \pm 0.10   $  & $13.00 \pm  0.12  $ & $12.84  \pm 0.17  $ \\
\hline
FeII$^{*}$    &  $\lambda$2333  & 1& $13.29 \pm 0.02$  & $12.90 \pm 0.02$ &    $13.37 \pm  0.02  $ & $13.36 \pm 0.02   $  & $12.40 \pm  0.06  $ & $12.30  \pm 0.05  $ \\
              &  $\lambda$2365  & 2& $12.66 \pm 0.05$  & $12.33 \pm 0.04$ &    $ < 12.2          $ & $ < 12.2          $  & $ < 12.2          $ & $ < 12.2          $ \\
              &  $\lambda$2389  & 3& $12.66 \pm 0.11$  & $ < 12.6       $ &    $ < 12.6          $ & $ < 12.6          $  & $ < 12.6          $ & $ < 12.6          $ \\
              &  $\lambda$2396  &2+3&$12.67 \pm 0.11$  & $12.15 \pm 0.12$ &    $12.13 \pm 0.12   $ & $ < 12.0          $  & $ < 12.0          $ & $ < 12.0          $ \\
\hline
FeII$^{**}$   &  $\lambda$2328  & 1& $13.03 \pm 0.01$  & $12.53 \pm 0.01$ &    $13.20 \pm  0.01  $ & $13.16 \pm 0.01   $  & $12.45 \pm  0.01  $ & $11.78  \pm 0.27  $ \\
              &                 & 2& $ < 13.0       $  & $ < 13.0       $ &    $ < 13.0          $ & $ < 13.0          $  & $ < 13.0          $ & $ < 13.0          $ \\
              &                 & 3& $ < 13.4       $  & $ < 13.4       $ &    $ < 13.4          $ & $ < 13.4          $  & $ < 13.4          $ & $ < 13.4          $ \\
              &                 &2+3&$ < 12.8       $  & $ < 12.8       $ &    $ < 12.8          $ & $ < 12.8          $  & $ < 12.8          $ & $ < 12.8          $ \\
\hline
FeII$^{***}$  &  $\lambda$2338  & 1& $12.86 \pm 0.02$  & $12.48 \pm 0.04$ &    $13.02 \pm  0.02  $ & $13.02 \pm 0.02   $  & $11.89 \pm  0.13  $ & $11.82  \pm 0.13  $ \\
              &  $\lambda$2359  & 2& $ < 13.0       $  & $ < 13.0       $ &    $ < 13.0          $ & $ < 13.0          $  & $ < 13.0          $ & $ < 13.0          $ \\
              &                 & 3& $ < 13.4       $  & $ < 13.4       $ &    $ < 13.4          $ & $ < 13.4          $  & $ < 13.4          $ & $ < 13.4          $ \\
              &                 &2+3&$ < 12.8       $  & $ < 12.8       $ &    $ < 12.8          $ & $ < 12.8          $  & $ < 12.8          $ & $ < 12.8          $ \\
\hline
FeII$^{****}$ &  $\lambda$2345  & 1& $12.54 \pm 0.02$  & $12.24 \pm 0.04$ &    $12.79 \pm  0.02  $ & $12.76 \pm 0.02   $  & $11.78 \pm  0.37  $ & $11.70  \pm 0.10  $ \\
              &  $\lambda$2414  & 2& $ < 12.7       $  & $ < 12.7       $ &    $ < 12.7          $ & $ < 12.7          $  & $ < 12.7          $ & $ < 12.7          $ \\
              &                 & 3& $ < 13.1       $  & $ < 13.1       $ &    $ < 13.1          $ & $ < 13.1          $  & $ < 13.1          $ & $ < 13.1          $ \\
              &                 &2+3&$ < 12.5       $  & $ < 12.5       $ &    $ < 12.5          $ & $ < 12.5          $  & $ < 12.5          $ & $ < 12.5          $ \\
\hline
FeII $^4$F    &  $\lambda$2332  & 1& $13.25 \pm 0.02$  & $12.18 \pm 0.24$ &    $13.62 \pm  0.01  $ & $13.42 \pm 0.02   $  & $12.37 \pm  0.12  $ & $12.12  \pm 0.23  $ \\
              &  $\lambda$2360  & 2& $ < 12.7       $  & $ < 12.7       $ &    $ < 12.7          $ & $ < 12.7          $  & $ < 12.7          $ & $ < 12.7          $ \\
              &                 & 3& $ < 13.1       $  & $ < 13.1       $ &    $ < 13.1          $ & $ < 13.1          $  & $ < 13.1          $ & $ < 13.1          $ \\
              &                 &2+3&$13.21 \pm 0.09$  & $12.6 \pm 0.36 $ &    $13.59 \pm 0.07   $ & $13.35 \pm 0.09   $  & $ < 12.5          $ & $12.37 \pm 0.52   $ \\
\hline
FeII$^4F^{*}$ &  $\lambda$2361  & 1& $12.73 \pm 0.04$  & $ < 11.5       $ &    $12.95 \pm  0.04  $ & $12.74 \pm 0.05   $  & $12.69 \pm  0.09  $ & $12.25  \pm 0.16  $ \\
              &                 & 2& $ < 12.7       $  & $ < 12.7       $ &    $ < 12.7          $ & $ < 12.7          $  & $ < 12.7          $ & $ < 12.7          $ \\
              &                 & 3& $ < 13.1       $  & $ < 13.1       $ &    $ < 13.1          $ & $ < 13.1          $  & $ < 13.1          $ & $ < 13.1          $ \\
              &                 &2+3&$ < 12.5       $  & $ < 12.5       $ &    $ < 12.5          $ & $ < 12.5          $  & $ < 12.5          $ & $ < 12.5          $ \\
\hline
FeII$^4D$     &  $\lambda$2563  & 1& $12.60 \pm 0.02$  & $11.72 \pm 0.16$ &    $11.99 \pm 0.10   $ & $11.80 \pm 0.15   $  & $ < 11.5          $ & $ < 11.5          $ \\
              &                 & 2& $ < 12.7       $  & $ < 12.7       $ &    $ < 12.7          $ & $ < 12.7          $  & $ < 12.7          $ & $ < 12.7          $ \\
              &                 & 3& $ < 13.1       $  & $ < 13.1       $ &    $ < 13.1          $ & $ < 13.1          $  & $ < 13.1          $ & $ < 13.1          $ \\
              &                 &2+3&$ < 12.5       $  & $ < 12.5       $ &    $ < 12.5          $ & $ < 12.5          $  & $ < 12.5          $ & $ < 12.5          $ \\
\hline
FeII$^4D^{*}$ &  $\lambda$2564  & 1& $12.37 \pm 0.05$  & $ < 11.5       $ &    $11.97 \pm 0.13   $ & $11.53 \pm 0.36   $  & $ < 11.5          $ & $ < 11.5          $ \\
              &                 & 2& $ < 12.7       $  & $ < 12.7       $ &    $ < 12.7          $ & $ < 12.7          $  & $ < 12.7          $ & $ < 12.7          $ \\
              &                 & 3& $ < 13.1       $  & $ < 13.1       $ &    $ < 13.1          $ & $ < 13.1          $  & $ < 13.1          $ & $ < 13.1          $ \\
              &                 &2+3&$ < 12.5       $  & $ < 12.5       $ &    $ < 12.5          $ & $ < 12.5          $  & $ < 12.5          $ & $ < 12.5          $ \\

\hline
\end{tabular}

All values are logarithmic cm$^{-2}$
}
\end{table*}

\begin{table*}[ht]
\caption{\bf The  \ion{Fe}{2} and \ion{Fe}{2}$^*$ column density ratios between
observation 1 and 2+3.}
\label{ratios}
\begin{tabular}{lccccccc}
\hline
\hline
     & I & II & III & IV & V & VI \\
\hline
\ion{Fe}{2}  &  $-0.35 \pm 0.05$ & $-0.13 \pm 0.12$  & $-0.35 \pm 0.10$  & $-0.21 \pm 0.12$ & $-0.24 \pm 0.14$& $-0.07 \pm 0.19 $ \\
\ion{Fe}{2}$^*$  &  $0.62  \pm 0.13$ & $0.75  \pm 0.14$  & $1.24  \pm 0.14 $ & $ > 1.36 $       & $ > 0.40 $      & $ > 0.30        $ \\

\hline
\end{tabular}

Ratios are expressed in logarithmic cm$^{-2}$
\end{table*}

\section{Distance of the absorbers from the GRB}

A constraint on the distance of the absorbing gas to the GRB can be
obtained using the ratio between the number of photons absorbed by the
first fine structure level of \ion{Fe}{2} and its corresponding ground
state. This ratio in the prompt spectrum of GRB080319B is 0.6 for
component I and II, between 0.3 and 0.4 for components III, IV, V and
VI.  Note that the value for component I and II is close to the
maximum theoretical value of 0.8. As a comparison, the same ratio in
the prompt spectrum of GRB060418 was 0.09 (Vreeswijk et al. 2007).
Calculations of population ratios (Silva \& Viegas 2002; see also
Prochaska, Chen \& Bloom 2006) show that the observed ratios are
obtained for a UV flux of $\sim 3\times 10^6 - 10^7\;G_0$ for the six
components, where $G_0=1.6\times 10^{-3}$ erg cm$^{-2}$ s$^{-1}$.
This implies distances from the GRB to the six absorbers
$R=\left[L_{UV}/(4\pi G_0\times (3\times
10^6-10^7))\right]^{1/2}\approx 18-34$ kpc (having assumed
$L_{UV}=6.7\times 10^{50}$ erg s$^{-1}$, obtained integrating the
light curve by Racusin et al. 2008b).

However, these population ratios are calculated assuming a
steady-state ionizing flux, an approximation which is not an
appropriate description for a GRB afterglow. To obtain a more reliable
result, we built up a time dependent photoexcitation code to compute
the column densities of the excited states as a function of the
absorbing gas distance from the GRB, in a similar way to that of
Vreeswijk et al. (2007). The basic equation to be
solved is the balance equation:

$$ {dN_u\over dt} = N_l B_{lu} F_{\nu}(\tau_0) - N_u[A_{ul} + B_{ul}F_{\nu}(\tau_0)], \eqno (1)$$

which describes the transition between two atomic levels. It gives the
increment in the upper level population $N_u$ as a function of the
lower level $N_l$, the flux $F_{\nu}(\tau_0)$ experienced by the
absorber, and the Einstein coefficients $A$ and $B$. In more detail,
$A_{ul}$ represents the spontaneous decay from the upper to the lower
state, $B_{ul}=A_{ul}\lambda^3/2hc $ the stimulated emission, and
$B_{lu}= B_{ul}g_u/g_l$ the absorption. Here $\lambda$ is the
transition wavelength and $g$ the degeneracy of the levels.
$F_{\nu}(\tau_0)$ is the monochromatic flux at the transition
frequency:

$$F_{\nu}(\tau_0) = F_{\nu}(0)e^{-\tau} + S_{\nu}(1-e^{-\tau}), \eqno (2)$$

corrected by the optical depth at the line center $\tau_0=
1.497\;10^{-2}N_l\lambda f / b$ (cgs units); $b$ is the Doppler factor
of the transition and $f$ its oscillator strength, which is related to the
Einstein coefficient $A$ by:

$$ f={m_e c A_{ul} g_u \lambda^2 \over 8 \pi^2 q_e^2 g_l}. \eqno (3) $$

The source function of the radiative transfer equation (2) is defined as:

$$ S_{\nu} ={ N_u(\nu)A_{ul}\over N_l(\nu)B_{lu} - N_u(\nu)B_{ul}} \eqno (4)$$

(Lequeux 2005). Finally, the uncorrected flux experienced by the absorber is:

$$F_{\nu}(0) = { F_{br} \; (t/t_{br})^{-\alpha_{br}} (\lambda/5439{\rm \AA})^{-\beta_{br}} (d_{L,GRB}/d)^2 \over 1+z}, \eqno (5)$$

(in cgs units) with z the GRB redshift used to compute its luminosity
distance $d_{L,GRB}$ and $d$ the distance of the absorber from the
GRB. The normalization constant $ F_{br}$ and the temporal and
spectral indices, $\alpha_{br}$ and $\beta_{br}$, have been taken from
the paper by Racusin et al. (2008b). The optical light curve of
GRB080319B in the V band (5439 \AA) is not monotonic, but can be
described by a broken power law with at least four different slopes in
the time interval between $20$ and $10^4$ s from the GRB. For each
break time $t_{br}$, we took the corresponding normalization constant
$ F_{br}$ and the temporal and spectral indices, $\alpha_{br}$ and
$\beta_{br}$, given in Racusin et al. (2008b).

Eq.1 must be simultaneously solved for many transitions, connecting in
principle all the levels of a given atom or ion (\ion{Fe}{2} in our
case). We included in our computation a total of 38 levels, the 16
lowest levels plus 22 higher excited states. The atomic data for the
transitions among these levels have been taken from Quinet et
al. (1996) (for transitions between the low energy states) and the
NIST database for other transitions (at the website
http://physics.nist.gov/PhysRefData/ASD/index.html). In order to
verify that the number of included transitions was large enough, we
ran our code with the input parameters used by Vreeswijk et al. (2007)
for GRB060418, and we found column densities fully consistent with
their results.

We stress that collisional processes and/or direct infrared pumping
(IR) alone can not be responsible for the variability we observe.  If
the first mechanism is at work, i.e. if the variability is produced by
a decreasing temperature, we should observe a reduction of all the
column densities of the excited states. Table 2 shows that fine
structure levels dramatically decrease, but the first excited level
(\ion{Fe}{2}$^4$F) stays almost constant in all components. On the
other hand, in case of pure IR pumping (assuming that the dominant UV
pumping process is for some reason inhibited), the fine structure
levels of the ground state should be more populated than those for
higher excited levels, which again is not observed. For more details
on the competition between such mechanisms, see again Vreeswijk et
al. (2007).

We ran our code using the total \ion{Fe}{2} column densities and
Doppler factors observed for components I and III ($N=1.16\;10^{14}$
and $ 1.88 \; 10^{14}$ cm$^{-2}$, $b=5$ and $10$ km s$^{-1}$,
respectively). The distance from the absorber was set as a free
parameter in order to obtain the best agreement between the data and
the photoexcitation code. In Fig.4 we show the results from our
code. Dotted, solid and dashed lines represent the predictions for
ground, fine structure and other excited levels, respectively. Short
(long) dashed lines are for \ion{Fe}{2} $^4$F and $^4$F* ($^4$D and
$^4$D*) levels. The data are reported as follows. Open circles
represent the ground state levels, closed circle the fine structures
of the ground state of \ion{Fe}{2}, open squares \ion{Fe}{2} $^4$F and
$^4$F* and open triangles \ion{Fe}{2} $^4$D and $^4$D*. The data
represent the first and second+third observation, and have been
slightly shifted to each other for clarity reasons. Fig. 4 shows that
the time evolution of the \ion{Fe}{2} column densities of component I
is best reproduced by a model with an absorber located at 2 kpc from
the GRB (lefthand plot), while the behaviour of component III is well
fitted with an absorber at 6 kpc from the GRB (righthand plot). The
closer the gas to the GRB, the longer the excited levels tend to be
populated with respect to the ground state. The ``anomalous''
behaviour of the \ion{Fe}{2} $^4$F level is due to its high
spontaneous decay rate toward the ground state, which is $\sim 3$
hours.

\begin{figure*}[ht]
\centering
\begin{tabular}{cc}
\includegraphics[width=8cm, height=8cm, angle=-0]{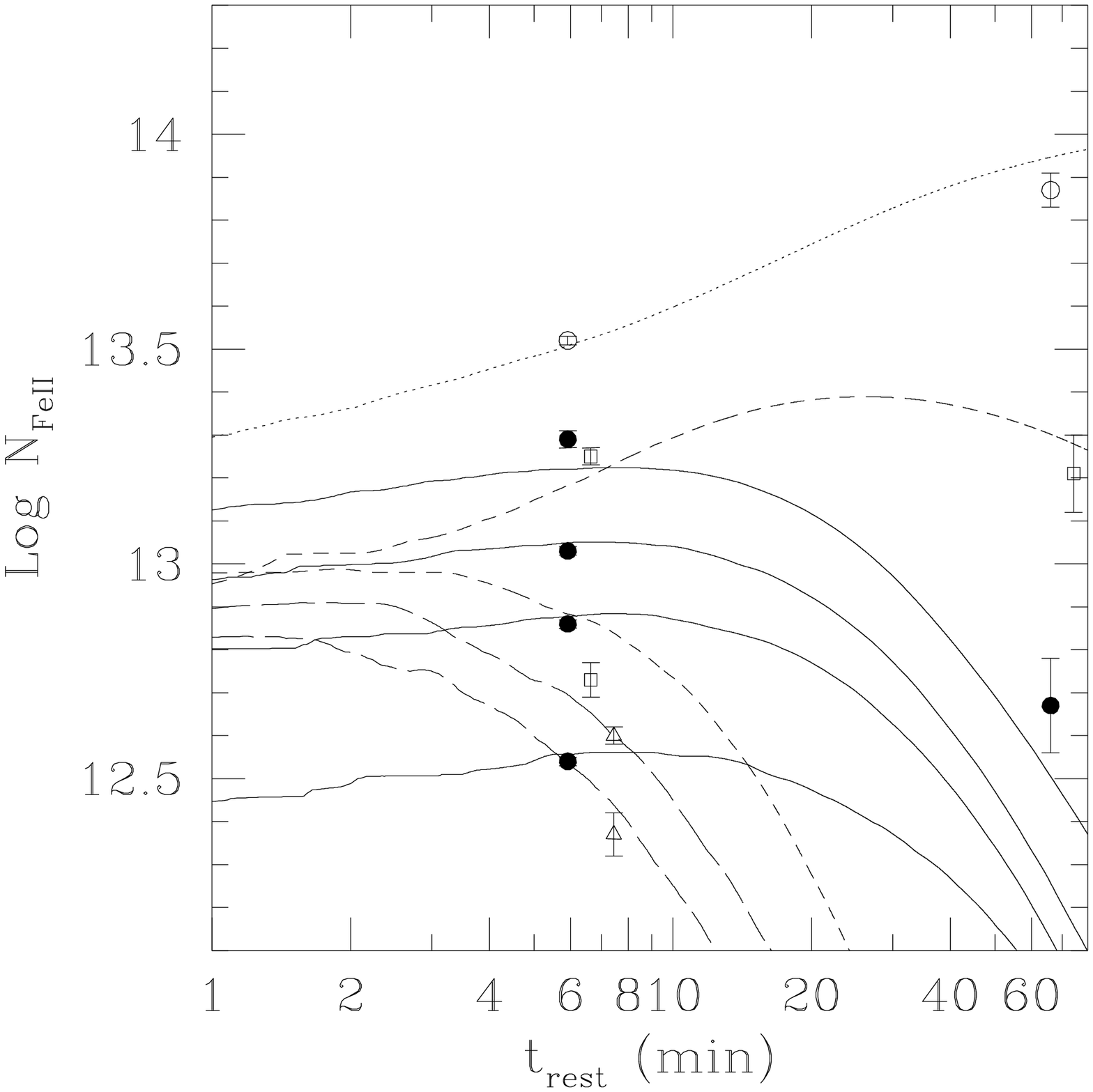}
\includegraphics[width=8cm, height=8cm, angle=-0]{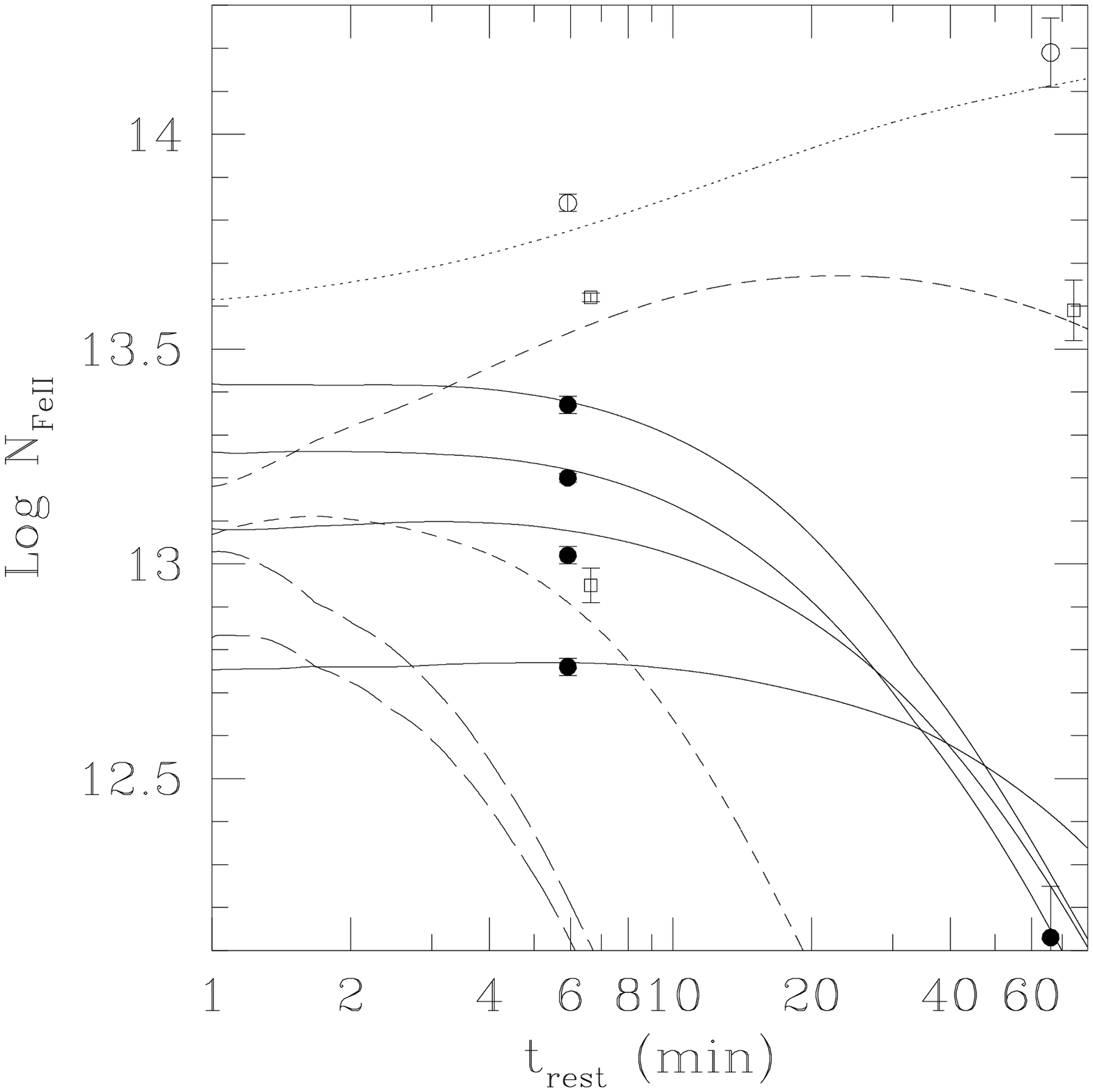}
\end{tabular}
\caption{Time evolution of \ion{Fe}{2} column densities for ground level
(open circles), fine structure level (solid circles) and first
(square) and second (triangles) excited level transitions for
component I (lefthand plot) and III (righthand plot) in the spectrum
of GRB080319b. Column density predictions from our time-dependent
photo-excitation code are also shown. They refer to the ground level
(dotted lines), fine structure levels (solid lines) and excited levels
(dashed lines) transitions, in the case of an absorber at 2 kpc
(lefthand plot) and at 6 kpc (righthand plot) from the GRB. For
clarity reasons, data points have been slightly shifted to each
others.}
\end{figure*}

In order for our results to be self-consistent, we need to make sure
that, at the smallest distance of 2 kpc as derived for component I,
\ion{Fe}{2} is not photoionized away by the strong UV radiation of the
burst. To this purpose, we performed a series of runs of the
photoionization code by Perna \& Lazzati (2002), which accounts for
the radiative-transfer of the radiation. We first simulated
a medium in thermal equilibrium at a temperature of $\sim 10^4$ K, and
let the radiation from the burst, modeled as in eq.(5), propagate
through it.  For a range of densities between $10^{-3}$ and $10^3$
cm$^{-3}$, we followed the concentration of \ion{Fe}{2} and
\ion{Mg}{1} absorbers at a distance of 2kpc, while the radiation from
the burst impinges on them. For densities $\sim 10^3$
cm$^{-3}$, the burst appears not to alter the initial concentration of
\ion{Fe}{2} and \ion{Mg}{1}. As the density decreases down to about
$10^{-2}$ cm$^{-3}$, the concentration of \ion{Fe}{2} still remains
unaltered, but \ion{Mg}{1} begins to be photoionized
significantly. This different behaviour is due to the fact that
\ion{Fe}{2} is screened by Hydrogen, because its photoionization
threshold is just above that of H. For even lower densities, \ion{Fe}{2}
begins to get photoionized away. For a density of $10^{-3}$
cm$^{-3}$, the concentration of \ion{Fe}{2} decreases by about 15\%
during the burst.  These calculations show that there is a wide range
of medium densities for which an \ion{Fe}{2} absorber at a distance of
2 kpc is not photoionized away by the radiation from the burst,
while, on the other hand, \ion{Mg}{1} is substantially
destroyed. Interestingly, component I is the only one for which
\ion{Mg}{1} is below the detection limit.

\section{Discussion and conclusions}

Thanks to the VLT RRM, which allowed the observation of GRB080319B in
just 5min (rest frame), we were able to catch the absorbing gas in a
highly excited state, producing the strongest \ion{Fe}{2} fine
structure lines ever observed in a GRB (or QSO) spectrum.  The optical
depth of these lines was dramatically reduced 2-3 hours later,
implying a factor of 4-20 decrease for all six components belonging to
the main absorption system. At the same time, the optical/UV flux
dropped by a factor of $\sim60$ (Bloom et al. 2008, Racusin et
al. 2008b). The variation of the \ion{Fe}{2} fine structure lines is
spectacular, when compared to previous GRB observations. Before
GRB080319B, the best case was certainly that of GRB060418 at z=1.490,
observed with UVES on comparably short timescales.  Vreeswijk et
al. (2007) report for this burst variations of the \ion{Fe}{2} fine
structure lines column densities by a factor of 1.4, in spectra taken
700 s and 7680 s after the GRB onset; in the same time interval the
optical/UV flux dropped by a factor of $\sim20$. The variations seen
in GRB080319B at similar rest frame timescales are clearly much more
prominent. This is probably due to the extremely intense optical/UV
radiation field of GRB080319B.

The optical GRB magnitude reached V$\sim5.3$ about 40 s after the start
of the GRB event.  At z=0.937, this magnitude implies a $\sim912$\AA{}
ionizing luminosity L$=1.2\times 10^{51}$ erg s$^{-1}$, assuming a
power law spectrum with frequency spectral index $-1$ and integrating
it up to 1 keV. Since the \ion{Fe}{2} ionization potential is just
above the photoionization edge of H, this ion is efficiently screened
and it can be photoionized only after H has been photoionized.  
%An order of magnitude estimate for the maximum H photoionization radius
%is given by $R=\sqrt{\frac{N_\gamma \sigma_H}{4\pi}}$ cm, where
%$N_\gamma$ is the number of photoionizing photons at 13.6 eV, and
%$\sigma_H=8\times 10^{-18}$ cm$^2$ is the H photoionization cross
%section.  
We can compute the number of ionizing photons by integrating
the optical/UV light curve (Bloom et al. 2008, Racusin et al. 2008b).
We find $N_\gamma= 8.6 \times 10^{62}$ ph at 912\AA; similar numbers
are obtained by extrapolating the XRT X-ray spectrum down to 912\AA{}
assuming no absorption, in addition to the Galactic value along the
line of sight.

%This yields an upper limit to the H photoionization
%radius $\ls7.5$kpc. At distances larger than this, H must be neutral,
%and \ion{Fe}{2} will not be fully photoionized.  Note that the maximum
%photoionization radius of \ion{Mg}{1} is comparable to that of H. With
%a photoionization cross section of $1.7\times 10^{-18}$ cm$^2$ and an
%ionization potential of 7.64 eV, the ionization radius of \ion{Mg}{1}
%is $\sim4.6$kpc. These estimates are confirmed by time evolving
%photoionization calculations using the observed GRB light-curve to
%estimate the input photoionization and assuming an optically thin
%medium (radiative transfer is neglected in this first approximation
%calculation). On the other hand, if radiative transfer in a medium of
%density $n_H$ were included, the upper limit to the photoionization
%radius would be given by $R={\frac{3\times N_\gamma}{4\pi n_H}}^{1/3}$
%cm, which gives $R>190 (n_H/cm^{-3})^{-1/3}$ pc.

We can constrain the distance of the absorbing gas to the GRB using
these numbers and the ratio between the number of photons absorbed by
the first fine structure level and the \ion{Fe}{2} ground state. In a
steady state approximation (Silva \& Viegas 2002; see also Prochaska,
Chen \& Bloom 2006), this distance turns out to be $\sim 18$ and $\sim
34$ kpc for component I and III, respectively. Since GRBs are highly
variable events, to refine these results, we built up a time dependent
photoexcitation code, to model the expected column densities of the
\ion{Fe}{2} levels as a function of time for an absorber illuminated
by a flux such as that of GRB080319B. We obtain smaller values for the
distances, namely, $\sim 2$ and $\sim 6$ kpc for component I and III,
respectively. This discrepancy can be explained by considering the
light curve of GRB080319B.  The flux of this GRB drops with a steep
power law (decay index $>5$) in the first 100 s (Racusin et
al. 2008b). The steady state approximation assumes a constant flux
from the GRB, with this constant being the total fluence radiated up
to the moment of the absorption line observation, divided by this time
range itself. Thus, this constant is $\sim 10^2$ times higher than the
real flux experienced by the absorber at the moment of the first UVES
observation. In this scenario, the steady state model will then
predict a larger distance in order to account for the higher fluxes at
later times.

To assure self-consistency, we need to make sure that, at the smallest
distance of 2 kpc as derived for component I, \ion{Fe}{2} is not
photoionized away by the strong UV radiation of the burst.  We showed
that there is a wide range of medium densities for which an
\ion{Fe}{2} absorber at a distance of 2 kpc is not photoionized away
by the radiation from the burst ($10^3 \div 10^{-2}$ cm$^{-3}$).On the
other hand, at densities below $\sim 1$ cm$^{-3}$, Mg I is
substantially destroyed. Interestingly, component I is the only one
for which \ion{Mg}{1} is below the detection limit.

Taken at face value, these distances are rather large for a typical
galaxy at z$\sim1$ (e.g. Sargent et al. 2007) and could imply that the
0.937 system is in the outskirts of the GRB host galaxy or in a nearby
clump along the line of sight.  Interestingly, HST imaging of the
field shows diffuse emission elongated south of the afterglow. In
particular, two faint clumps of emissions are located at $1.5''$ and
$3''$ from the afterglow (Tanvir et al. 2008). At z=0.937 these
correspond to projected distances of 12 and 24kpc, and may suggest the
presence of a complex structure of clumps around the GRB host
galaxy. If this is the case, the absorbers may well belong to one of
these clumps.

%Component I has the highest ratio of photons absorbed by \ion{Fe}{2}
%excited and ground states, suggesting again that this component is the
%closest one to the GRB site.  Note that this component shows
%\ion{Fe}{2} fine structure line variations significantly smaller than
%the other components (viz. III and IV, Fig. \ref{var} and Table
%3). The decay time of the first excited state ($\sim 500$ s, Vreeswijk
%et al. 2007) is shorter than or comparable to the rest frame time
%between the first and second epoch spectra. This implies that it is
%easier to maintain a high level of excitation for a longer time for an
%absorber closer to the source of UV photons than for farther
%absorbers, in agreement with our previous results. Alternatively,
%collisional excitation may help in maintaining a sizeable population
%of excited levels in this component.  Component I does not show strong
%\ion{Mg}{1} absorption, unlike all other components.  Both analytical
%and time evolving photoionization calculations show that \ion{Mg}{1}
%should be present at distances $>10$ kpc, where the analysis of the
%fine structure lines suggest that also component I is located.  The
%lack of \ion{Mg}{1} in component I can be explained by a gas
%temperature $>5\times10^4$K (Shull \& Van Steenberg 1982), while the
%temperature of the gas of the other components should be lower than
%this value. This is a further confirmation that the ISM of high-z
%galaxies is complex, structured and characterized by numerous distinct
%components with different temperature and (most likely) density.

\acknowledgments
We acknowledge support from ASI/INAF contracts
ASI/I/R/039/04 and ASI/I/R/023/05/0. SDV is supported by SFI.

\end{document}